# Electronic properties of a π-conjugated Cairo pentagonal lattice: Direct band gap, ultrahigh carrier mobility and slant Dirac cones

Xiaofei Shao[1], Xiaobiao Liu[1], Xinrui Zhao[2], Junru Wang[1], Xiaoming Zhang[1], Mingwen Zhao[1*]

[1] *School of Physics and State Key Laboratory of Crystal Materials, Shandong University, Jinan 250100, Shandong, China*

[2] *School of the Gifted Young, University of Science and Technology of China, Hefei, 230026, Anhui, China.*

**Corresponding Author**

*zmw@sdu.edu.cn

**KEYWORDS:** P-d π-conjugation; Direct band gap; Ultra-high carrier mobility; 2D pentagonal materials; Tight-binding model; First-principles calculations;

# ABSTRACT

Two-dimensional (2D) lattices composed exclusively of pentagons represent an exceptional structure of materials correlated to the famous pentagonal tiling problem in mathematics, but their π-conjugation and the related electronic properties have never been reported. Here, we propose a tight-binding (TB) model for a 2D Cairo pentagonal lattice and demonstrate that *p-d* π-conjugation in the unique framework leads to intriguing properties, such as an intrinsic direct band gap, ultra-high carrier mobility and even slant Dirac cones. On the basis of first-principles calculations, we predict a candidate material, 2D penta-$NiP_2$ monolayer, derived from bulk $NiP_2$ crystal, to realize the predictions of the TB model. It has ultra-high carrier mobility ($\sim 10^5$-$10^6$ $cm^2 V^{-1} s^{-1}$) comparable to that of graphene and an intrinsic direct band gap of 0.818 eV, which are long desired for high-speed electronic devices. The stability and possible synthetic routes of penta-$NiP_2$ monolayer are also discussed.

## I. INTRODUCTION

The π-conjugation of $p_z$ orbitals in graphene, a two-dimensional (2D) honeycomb (hexagon) lattice, leads to the unique linear energy-momentum dispersion relation (Dirac cone) [1,2] and extremely high carrier mobility that can attain $10^5$-$10^7$ cm$^2$V$^{-1}$s$^{-1}$[3-5]. However, pristine graphene does not possess a band gap, a property that is crucial for application in logic transistors to attain a large on/off ratio. An opportune band gap and ultra-high carrier mobility are thus highly desirable for the relevant applications.

Due to the restriction of translation symmetry, fivefold-rotation symmetry is prohibited in lattice structures, which excludes regular pentagon from the units of 2D lattices. Nevertheless, there exist convex pentagons that can tile the Euclid plane monohedrally (i.e. with one type of tile), known as the famous pentagonal tiling problem in mathematics. Fifteen types of such convex pentagons have been found up to now [6], offering mathematical basis of 2D pentagonal materials. Notably, isohedral pentagonal Cairo with edge-to-edge features represents a high symmetry case of the 15 monohedral tilings above, which are promising for design of 2D pentagonal materials. Indeed, pentagonal graphene (namely penta-graphene) derived from the Cairo tiling has been proposed theoretically [7]. The sp$^3$-hybridized carbon atoms in the buckled penta-graphene break the π-conjugation of the *$p_z$* orbitals of *sp$^2$* carbon atoms, resulting in an indirect band gap of 3.25 eV [7], which is inactive for optoelectronic applications. Following penta-graphene, a series of pentagonal 2D materials with similar lattice structures were proposed theoretically [8-14]. However, neither π-conjugation features nor direct band gap is found in these 2D penta-materials and none of them has been synthesized in experiments. As the first real 2D penta-material, PdSe$_2$ few-layers with puckered morphology was recently realized by exfoliating bulk crystal [15] .Unfortunately, it remains an indirect band gap semiconductor (~1.3 eV) with carrier mobility of only about $10^2$ $cm^2V^{-1}s^{-1}$.

In this work, we propose a tight-binding (TB) model of $d_\pi$ and $p_z$ orbitals in a 2D Cairo pentagonal lattice and demonstrate that the *p-d* π-conjugation in the unique framework can lead to long-desired properties: an intrinsic direct band gap, ultra-high carrier mobility and even slant Dirac cones, that have never been reported in 2D penta-materials. On the basis of density-functional-theory (DFT) calculations, we propose a candidate material to realize the predictions of the TB model: penta-NiP$_2$ monolayer derived from the bulk crystal. The intrinsic direct band gap of penta-NiP$_2$ monolayer is predicted to be 0.818 eV and the theoretical carrier mobility evaluated from an acoustic phonon-limited scattering model can reach ~$10^5$ $cm^2V^{-1}s^{-1}$ which is much higher than that of black phosphorene [16] and even comparable to that of graphene [17]. The stability and plausibility of penta-NiP$_2$ monolayer are confirmed by the energetic favorability, phonon spectrum, molecular dynamics simulations, and etc. The π-conjugated Cairo pentagonal lattice with diverse electronic band structures demonstrated by both TB model and DFT calculations enriches the family of 2D pentagonal materials and may find applications in high-speed electronic devices.

## II. METHODS

Our first-principles calculations were performed within density-functional theory (DFT) by using the Vienna *ab initio* simulation package (VASP) [18-20]. The energy cutoff of the plane-wave basis set [21] was 600 eV. The electron-ion interaction was described by a projector augmented wave (PAW) approach [22, 23]. A generalized gradient approximation (GGA) [24] in the form of Perdew-Burke-Ernzerhof (PBE) was employed for the structural optimization. Van der Waals interactions were considered using the zero damping DFT-D3 method [25] for penta-NiP$_2$ bilayer and bulk crystal. For penta-NiP$_2$ monolayer, the unit cell was repeated periodically along the x- and y- directions, while a large vacuum space of 30 Å was applied along the z-direction to avoid interaction between adjacent images. The Brillouin zone (BZ) was sampled by using 7×7×1 *k*-point mesh in structural optimization according to the Monkhorst-Pack (MP) method [26]. All the atoms were fully relaxed without any symmetry restriction until the residual forces on each atom are smaller than $10^{-3}$ eV/Å. The criterion for energy convergence is $10^{-8}$ eV/cell. To avoid the self-interaction errors of the PBE functional, a hybrid functional in the form of Heyd-Scuseria-Ernzerhof (HSE06) [27] was also employed in calculations of electronic band structures and carrier mobility. The spin-orbit coupling (SOC) effect was not taken into account. The phonon spectrum was calculated by using the Phonopy code [28] interfaced with VASP. The *ab initio* molecular dynamics simulations (AIMDS) were performed in an NVT ensemble at temperature of 500K with a time step of 0.5 *fs*.

The carrier mobility of penta-NiP$_2$ monolayer was calculated using an acoustic phonon-limited scattering model. In this model, the optical phonon scattering mechanism which may have comparable effect on mobility as the acoustic ones [29] was not taken into account. For a 2D material, the carrier mobility $\mu_{2D}$ is determined by the effective mass in the transport direction ($m_x^*$), average effective mass ($m_d$), deformation potential ($E_1$), elastic modulus ($C_{2D}$) and temperature (T) in the expression [30-32]

$$\mu_{2D} = \frac{e\hbar^3 C_{2D}}{k_B T m_e^* m_d (E_1^i)^2} \quad \ldots (1)$$

The average effective mass is defined as $m_d=(m_x^* m_y^*)^{1/2}$, where $m_y^*$ is the effective mass of charge carriers (electron and hole) along the direction perpendicular to the transport direction. The effective mass is calculated by fitting the band line in the region near the Fermi level. The deformation potential constant $E_1$ of the VBM for hole or CBM for electron along the transport direction is $E_1^i = \Delta V_i/(\Delta l/l_0)$, where $\Delta V_i$ represents the energy change of $i^{th}$ band evaluated with respect to the vacuum level under proper cell compression and dilatation, $l_0$ is the lattice constant in the transport direction and $\Delta l$ is the deformation of $l_0$. The elastic modulus $C_{2D}$ of the longitudinal strain in the propagation directions of the longitudinal acoustic wave is derived from $(E-E_0)/S_0 = C(\Delta l/l_0)^2/2$, where $E$ is the total energy and $S_0$ is the lattice volume at equilibrium for a 2D system. The temperature used for the mobility calculations was 300 K. Notably, the $C_4$ symmetry of penta-NiP$_2$ monolayer leads to isotropic elastic constants, effective masses and carrier mobility along two perpendicular orientations. This differs from the cases of anisotropic 2D materials like black phosphorene, where the carrier mobility will be largely overestimated without anisotropic effect correction [32]. In this work, we evaluated the carrier mobilities along the [100] and [110] directions of penta-NiP$_2$ monolayer.

### III. RESULTS AND DISCUSSION

1. **Tight-binding model.**

The 2D Cairo pentagonal lattice considered in this work is shown in Fig. 1(a), which belongs to a square Bravais lattice. It represents a special case of type 4 tiling, one of fifteen known pentagons to tile the plane monohedrally. In this tiling pattern, the pentagon has five inner angles of 135°-$\theta$, 135°-$\theta$, 90°, 90°+2$\theta$ and 90° and two types of sides. The Cartesian coordinates in the unit of $a_0$ of the six sites in a primitive cell are (0,0), ($\alpha + \beta, \alpha + \beta$), ($\beta$, -$\alpha$), (-$\alpha$, -$\beta$), (-$\beta$, $\alpha$), and ($\alpha$, $\beta$), where $\alpha=sin\theta$ and $\beta=cos\theta$. $\theta$ represents the angle between a lattice vector and a side connecting a threefold-coordinated vertex and a fourfold-coordinated vertex and $a_0$ is the length of the side, as denoted in Fig. 1(a). The length of the basis vector is $a = 2a_0(\alpha + \beta)$. Variation of $\theta$ doesn't change the symmetry of the lattice.

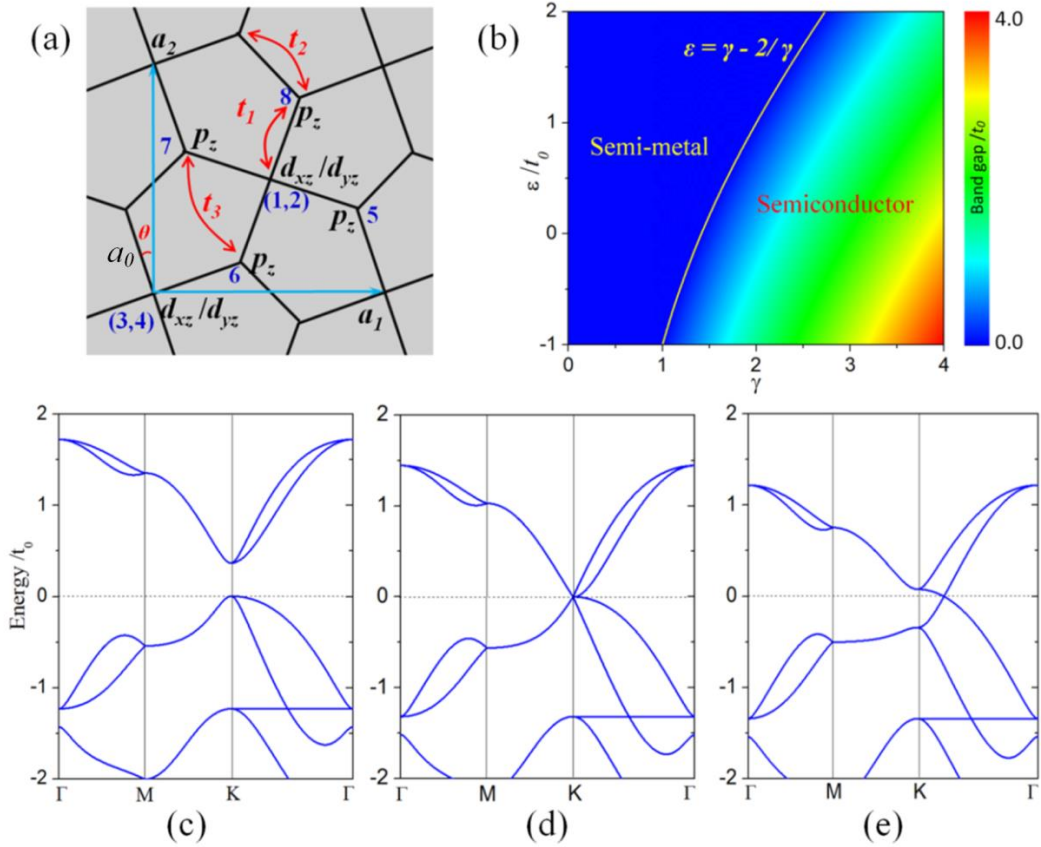

**Figure 1** (a) Schematic representation of the tight-binding (TB) model involving nearest ($t_1$ and $t_2$) and next-nearest ($t_3$) hoppings. Atomic orbitals in the unit cell are numbered by the blue numbers. $\boldsymbol{a_1}$ and $\boldsymbol{a_2}$ are the two basis vectors which are taken as *x*- and *y*-directions. (b) The energy band diagram in the parameter space ($\varepsilon/t_0$, $\gamma$) of the TB model. The yellow line indicates the boundary between semi-metals and semiconductors. (c)-(e) The TB band structures of (c) semiconducting phase with $\varepsilon < (\gamma - 2/\gamma)t_0$ and semi-metallic phases with (d) $\varepsilon = (\gamma - 2/\gamma)t_0$ and (e) $\varepsilon > (\gamma - 2/\gamma)t_0$, respectively. The coordinates of the highly-symmetric points in the square BZ are $\Gamma$ (0,0), M($\pi/a$, 0) and K($\pi/a$, $\pi/a$). The energy at the valence band maximum is set to zero

We assume that each primitive cell of the lattice contains eight atomic orbitals. Two orthorhombic *d* orbitals ($d_{xz}$ and $d_{yz}$) reside at the fourfold-coordinated site, while the threefold-coordinated site accommodates a single $p_z$ orbital. The onsite energy difference between the *d* and *p* orbitals is represented by $\varepsilon$. For simplification, we only considered the electron hoppings between nearest-neighbor (NN) sites and next-nearest-neighbor (NNN) sites. The *p-d* hopping between NN sites has amplitude of $t_1$, while the *p-p* hoppings between NN sites and between NNN sites have the amplitudes of $t_2$ and $t_3$, respectively, as denoted in Fig. 1(a). For a planar configuration, the *p-d* hopping energy ($t_{ij}$) between site *i* and *j* is determined by the vector ($\vec{r}_{ij}$) from *i* to *j*, according to the expressions, $t_{i,j}(p_z \to d_{xz}) = lV_{pd\pi}$, and $t_{i,j}(p_z \to d_{yz}) = mV_{pd\pi}$, where *l* and *m* represent the cosines of the angles between $\vec{r}_{ij}$ and the *x*-, *y*- or *z*- direction. Setting $a_0 = 1$, $V_{pd\pi} = -t_0$ ($t_0 > 0$), $t_1 = -\gamma t_0$ and $t_2 = -\gamma' t_0$, the TB Hamiltonian of the Cairo pentagon tiling can be written as:

$$H(\vec{k}) = -t_0 \begin{pmatrix} -\varepsilon/t_0 & 0 & 0 & 0 & \beta S_{15} & -\alpha S_{16} & -\beta S_{15}^* & \alpha S_{16}^* \\ & -\varepsilon/t_0 & 0 & 0 & \alpha S_{15} & \beta S_{16} & -\alpha S_{15}^* & -\beta S_{16}^* \\ & & -\varepsilon/t_0 & 0 & \alpha S_{35} & -\beta S_{36} & -\alpha S_{35}^* & \beta S_{36}^* \\ & & & -\varepsilon/t_0 & -\beta S_{35} & -\alpha S_{36} & \beta S_{35}^* & \alpha S_{36}^* \\ & & & & 0 & \gamma' S_{56} & \gamma S_{57} & \gamma' S_{58} \\ & & & & & 0 & \gamma' S_{58} & \gamma S_{68} \\ & & & & & & 0 & \gamma' S_{56}^* \\ & & & & & & & 0 \end{pmatrix} \quad \ldots (2)$$

where $S_{15} = e^{i(\beta k_x - \alpha k_y)}$, $S_{16} = e^{-i(\alpha k_x + \beta k_y)}$, $S_{35} = e^{i(-\alpha k_x + \beta k_y)}$, $S_{36} = e^{i(\beta k_x + \alpha k_y)}$, $S_{57} = e^{i(2\alpha k_x + 2\alpha k_y)}$, $S_{68} = e^{i(2\alpha k_x - 2\alpha k_y)}$ $S_{56} = e^{i(-(\alpha+\beta)k_x + (\alpha-\beta)k_y)} + e^{i((\alpha+\beta)k_x + (\alpha-\beta)k_y)}$, $S_{58} = e^{i((\alpha-\beta)k_x + (\alpha+\beta)k_y)} + e^{i((\alpha-\beta)k_x - (\alpha+\beta)k_y)}$. We give only half of the matrix for simplification. The whole matrix is filled to ensure a Hermitian matrix.

By diagonalizing the TB Hamiltonian, we obtained eight bands in the reciprocal space which are highly dependent on the hopping parameters ($t_0$, $\gamma$ and $\gamma'$) and onsite energy ($\varepsilon$), as shown in Fig. 1(c)-(e). The TB Hamiltonian of the pentagonal lattice gives an intrinsic band gap ($E_g$) at the K ($\pi/a$, $\pi/a$) point (the corner of the square BZ) between band 6 and band 7 with

$$E_g = \left((3\gamma t_0 - \varepsilon) - \sqrt{(\varepsilon + \gamma t_0)^2 + 16 t_0^2}\right)/2. \quad \ldots(3)$$

It is interesting to see that the band gap value is independent of the angle θ and $\gamma'$. The profile of the band lines is also independent of the angle $\theta$, because variation of $\theta$ doesn't change the symmetry of the lattice. The band gap diagram in the parameter space ($\varepsilon/t_0$, $\gamma$) is plotted in Fig. 1(b). The two phases, semiconductor and semimetal, are separated by the yellow line which satisfies $\varepsilon = (\gamma - 2/\gamma)t_0$. As $\varepsilon < (\gamma - 2/\gamma)t_0$, $E_g > 0$ suggests a semiconducting solution, as shown in Fig. 1(c). In the region near the K point, bands 6 and 7 are rather dispersive, suggesting small effective masses and thus high carrier mobility of electrons and holes. For $\varepsilon = (\gamma - 2/\gamma)t_0$ the valence and conduction bands touch at a point, where the electronic states are fourfold degenerate, as shown in Fig. 1(d). The coexistence of linear dispersion relation (massless Dirac fermions) along the K-M direction and parabolic dispersion relation (mass fermions) along the K-Γ direction may bring about new phenomena distinct from pure massless Dirac fermions. For $\varepsilon > (\gamma - 2/\gamma)t_0$, band inversion between bands 6 and 7 takes place, leading to the Dirac cone residing between Γ and K points, as shown in Fig. 1(d). Due to the $C_4$ symmetry of the lattice, there are four Dirac cones in the Brillouin zone. It is noteworthy that this pentagonal framework has slant Dirac cones with anisotropic Fermi velocities distinct from graphene with isotropic Dirac cones. This is the first prediction on the existence of Dirac cones in 2D pentagonal materials, enriching the 2D Dirac material family. The diversity of the electronic band structures of the *p-d* π-conjugated Cairo pentagonal lattice opens an avenue for design of 2D pentagonal materials with desired electronic properties.

## 2. Electronic band structure of penta-NiP$_2$ monolayer.

To realize the electronic band structures predicted by the TB model, we propose a 2D material with a Cairo pentagonal lattice, namely penta-NiP$_2$ monolayer. It is built by placing Ni atoms on the fourfold-coordinated sites and P atoms on the threefold-coordinated sites, as shown in Fig. 2(a). After structural optimization, penta-NiP$_2$ monolayer exhibits planar configuration with a space group of P4g. In contrast to the 2D pentagonal materials proposed in the previous works [8-15], it shows a perfect Cairo pentagonal titling pattern without any buckling. We attribute it to the Ni(II) ions with $d^8$ electron configuration which prefer square planar molecular geometries. For penta-NiP$_2$ monolayer, $\theta$ is determined to be about 20° and the Ni-P and P-P bonds (the two types of sides of the pentagon) have the lengths of 2.167 Å and 2.091 Å, respectively. The bond lengths are very close to those of monoclinic NiP$_2$ crystal (2.196 Å and 2.225 Å). The structure parameters and comparison with the bulk crystals [33-35] are listed in Table I. The stability and plausibility of penta-NiP$_2$ monolayer will be discussed in the last section.

We then explored the electronic properties of penta-NiP$_2$ monolayer from DFT calculations. The electronic band structures calculated by using PBE and HSE06 functionals are plotted in Fig. 2(b) and 2(c), respectively. Both functionals give a direct band gap at the K ($\pi/a$, $\pi/a$) point, in good agreement with the TB model, but the band gap values strongly depend on the employed functional, which are 0.069 eV (PBE) and 0.818 eV (HSE06), respectively. The band gap is smaller than that of silicon (1.1-1.4 eV) [36] and black phosphorene (1.5 eV) [37] and is thus promising for long wavelength optoelectronic applications.

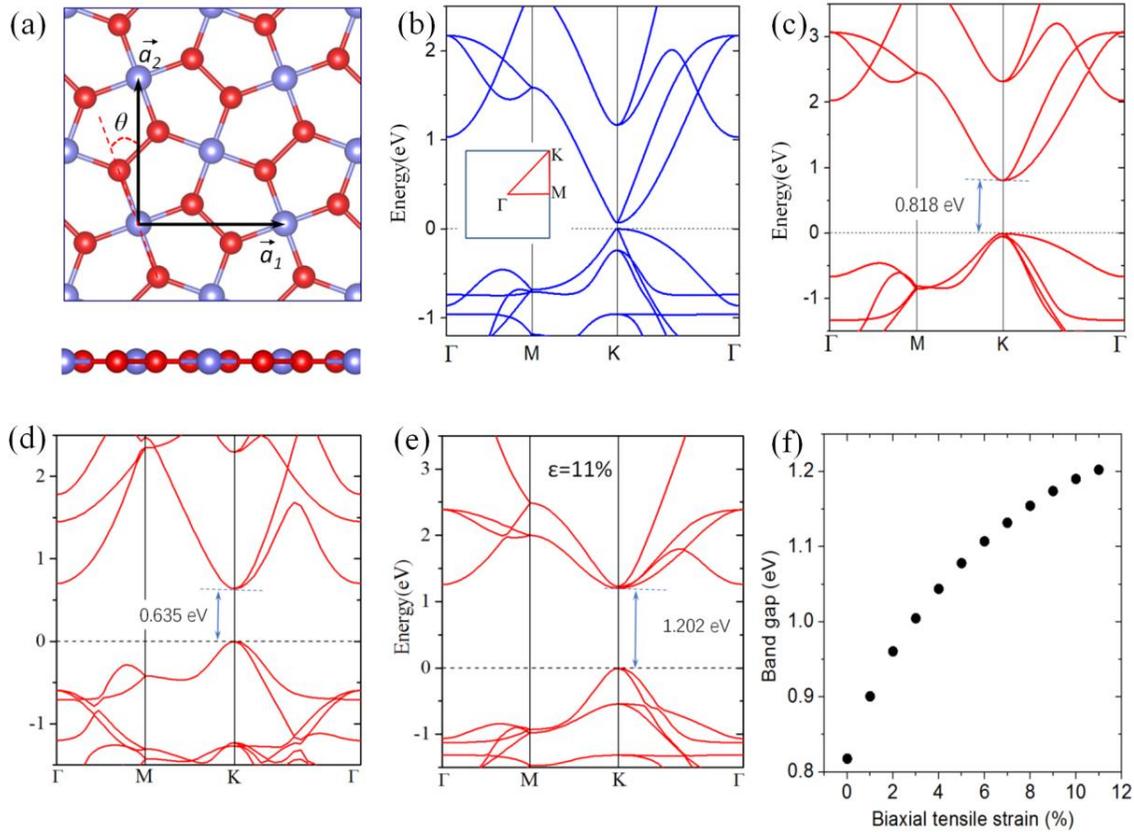

**Figure 2**. (a) Top and side views of the atomic structure of 2D penta-$NiP_2$ monolayer. (b)-(e) Electronic band structures of penta-$NiP_2$ monolayer obtained by using (b) PBE and (c) HSE06 functionals, (d) penta-$NiP_2$ bilayer, (e) penta-$NiP_2$ monolayer under tensile strain of 11%. The energy at the VBM is set to zero. (f) Variation of band gap of penta-$NiP_2$ monolayer as a functional tensile strain obtained by using HSE06 functional.

For the AA-stacked penta-$NiP_2$ bilayer, band gap reduces to 0.635 eV due to the interlayer interaction, but the direct band gap remains intact, as shown in Fig. 2(d). The robust direct band gap in penta-$NiP_2$ bilayer is superior to that in $Tl_2O$ bilayer, where the interlayer interaction takes it to an indirect band gap semiconductor [38]. The direct band gap of penta-$NiP_2$ monolayer is also preserved under uniform tensile strain, as shown in Fig. 2(e). With the increase of tensile strain, the band gap increases gradually, as shown in Fig. 2(f). At a tensile strain of 11%, the band gap attains 1.202 eV, which is about 1.5 times larger than that of the equilibrium state. The robust direct band gap of penta-NiP$_2$ confirms that it is the intrinsic feature of the lattice protected by the lattice symmetry. The strain-modulated energy band gap is promising for mechanical-electronic devices.

The origins of the dispersive bands near the Fermi level can be revealed by the orbital-resolved electron density of states. We projected the electron density of states (PDOS) onto the atomic orbitals of Ni and P atoms, as shown in Fig, 3(a) and 3(b), respectively. Clearly, the electronic states in the proximity of the Fermi level arise mainly from the $d_{xz}/d_{yz}$ orbitals of Ni atoms and the $p_z$ orbitals of P atoms. According to the crystal field theory [39,40], in a crystal-field environment with a square symmetry, the 3d orbitals of the Ni ions split into $d_{z^2}$, $d_{xy}$, $d_{x^2-y^2}$ orbitals and a double-degenerate $d_\pi$ ($d_{xz}$ +$d_{yz}$) orbitals. The inequivalence of the $d_{xz}$ and $d_{yz}$ orbitals in the PDOS is attributed to the angle (θ = 20°) between the Ni-P bond and x- (or y-) direction, as shown in Fig. 2(a). Notably, there is overlap in energy between the $d_{xz}/d_{yz}$ orbitals of Ni atoms and the $p_z$ orbital of P atoms near the Fermi level, implying the strong coupling between them. The π-π interaction between the $p_z$ and $d_\pi$ orbitals leads to π-conjugation throughout the pentagonal framework, which is responsible for the dispersive bands in the proximity of the Fermi level. These features are further confirmed by the charge density of the electron wavefunctions of the VBM and CBM, as shown in Fig. 3(c) and 3(d), respectively. The isosurfaces of these electron wavefunctions exhibit π-π binding (for VBM) and anti-binding (for CBM) characteristics between the $p_z$ and $d_\pi$ orbitals. The orbital contribution to the bands is consistent with the TB model.

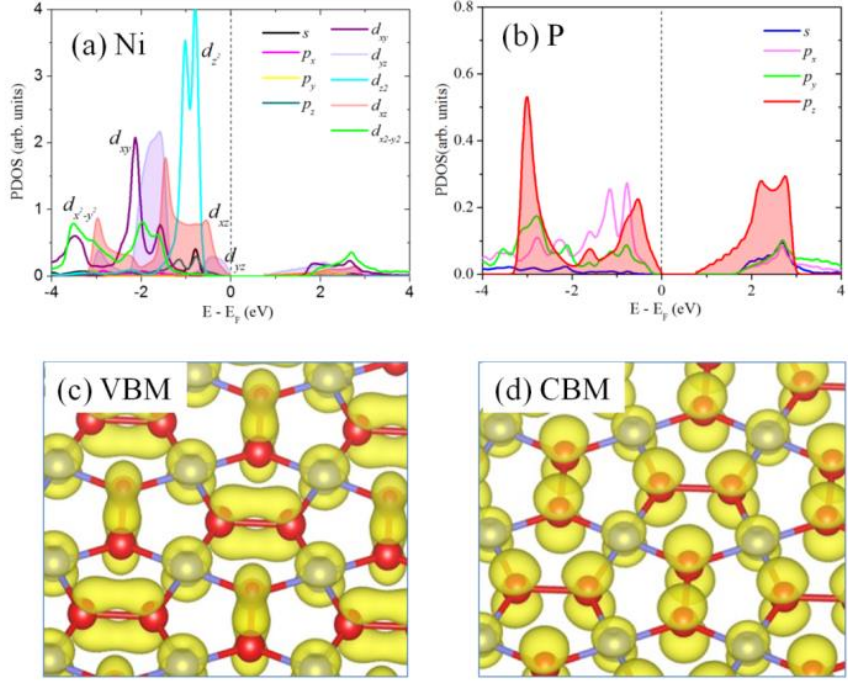

**Figure 3** Orbital-resolved electron density of states (PDOS) projected onto (a) Ni(II) and (b) P atoms, respectively. The energy at the VBM is set to zero. (c) The isosurfaces of the Kohn-Sham electron wavefunctions of (c) valence band maximum (VBM) (d) and conduction band minimum (CBM).

The Ni ion in the penta-NiP$_2$ monolayer is at Ni(II) state with electron occupation of d$^8$ ($d_{xy}^{\uparrow\downarrow} d_{z^2}^{\uparrow\downarrow} d_{xz}^{\uparrow\downarrow} d_{yz}^{\uparrow\downarrow}$), while the P$^-$ ion has one unpaired $p_z$ electron, as shown in Fig. 4(a). Therefore, each unit cell has twelve electrons (eight from the $d_{xz}$ and $d_{yz}$ orbitals of Ni and four from the $p_z$ orbitals of P atoms). According to the TB model, the Fermi level resides right within the band gap. The direct band gap and dispersive bands near the Fermi level are in agreement with the TB model. The valence and conduction bands nearest to the Fermi level obtained from the DFT-HSE06 method can be well fitted by the TB model using the parameters, $t_0 = 1.7$ eV, $\gamma = 1.6$, $\gamma' =$ 0.6, $\varepsilon = 0.34$ eV, and the structural parameters of penta-NiP$_2$ monolayer taken from DFT calculations, as shown in Fig. 4(b), which confirms the *p-d* π-conjugation features of penta-NiP$_2$ monolayer. It is noteworthy that the highest valence band and the lowest conduction band exhibit high anisotropy in the reciprocal space, as shown in Fig. 4(c), which will result in anisotropic electron transportation.

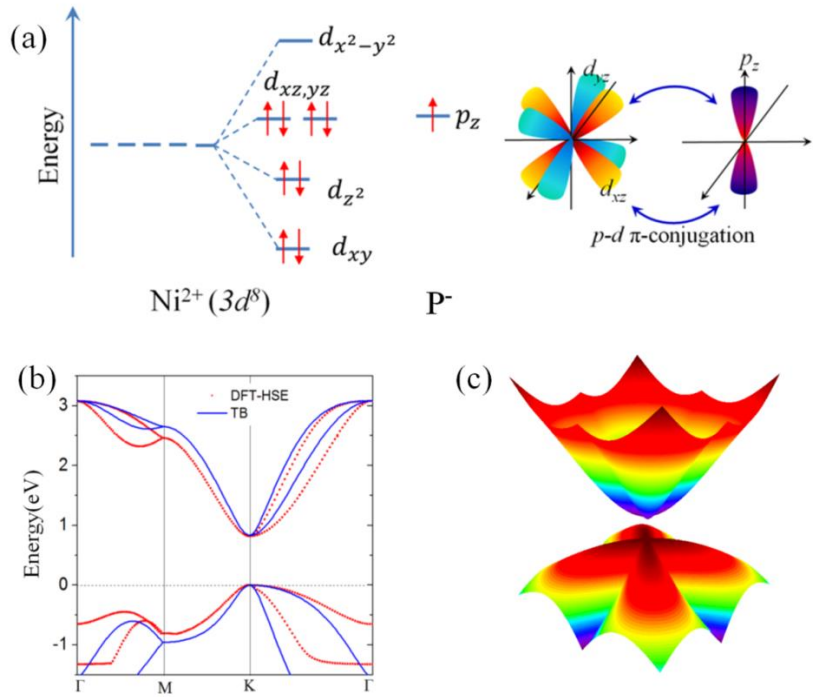

**Figure 4** (a) Schematic representation of the *d*-orbital splitting and *p-d* π-conjugation. (b) Electronic band structures TB model (blue line) with the parameters of $t_0 = 1.7$ eV, $\gamma = 1.6$, $\gamma' = 0.6$, and $\varepsilon = 0.34$ eV. The band lines of penta-NiP$_2$ monolayer nearest to the Fermi level obtained from DFT-HSE06 strategy are indicated by the red dots. The energy at the VBM is set to zero. (c) Three-dimensional plot of the highest valence band and the lowest conduction band near the Fermi level in the reciprocal space.

In addition, from the electronic band structures of penta-NiP$_2$ monolayer, we can deduce that the diverse electronic band structures predicted by the TB model may be realized in penta-MX$_2$ monolayer with M = Ni, Pt and Pd, X = N, P, As and Sb. The electronic band structures of penta-MX$_2$ materials obtained by using DFT calculations are plotted in Fig. S1 of the Supplementary Information. It is interesting to see that the semiconducting or semi-metallic band structures of these penta-MX$_2$ materials are consistent with the predictions of the TB model, revealing a new family of 2D pentagonal materials

## *3. Carrier mobility of penta-NiP$_2$ monolayer*

The dispersive bands near the Fermi level suggest that high carrier (electron and hole) mobility may be achieved in penta-NiP$_2$ monolayer. Using an acoustic phonon-limited scattering model, we evaluated the carrier mobility of penta-NiP$_2$ monolayer along [100] and [110] directions, respectively. The present calculations give only the upper bound of the carrier mobility, because other important scattering mechanisms, such as optical phonons, Coulomb and defects, are not included, which may contribute comparably to the acoustic phonon scattering. The carrier mobilities and the relevant parameters obtained by using PBE and HSE06 functionals are listed in Table II. PBE functional predicts ultra-high hole mobility of about $1 \times 10^6$ cm$^2$V$^{-1}$s$^{-1}$ along the [100] direction, even higher than the theoretical value of graphene ($3 \times 10^5$ cm$^2$V$^{-1}$s$^{-1}$) evaluated using the same theoretical strategy [17]. The electron mobility is also as high as $2 \times 10^5$ cm$^2$V$^{-1}$s$^{-1}$. Such ultra-high carrier mobilities arise mainly from the small effective masses of electron and hole, ~0.03 m$_0$, $m_0$ is the mass of a free electron. HSE06 functional gives not only a larger band gap but also larger effective masses of electron and hole, because the band dispersion is altered, especially for the valence bands. Even though, the electron mobility along the [100] direction remains as high as $0.4 \times 10^5$ cm$^2$V$^{-1}$s$^{-1}$, which is four times higher than the highest value in black phosphorus [16]. The hole mobility of penta-NiP$_2$ monolayer along the [100] direction is about 2×10$^3$ cm$^2$V$^{-1}$s$^{-1}$. Notably, the carrier mobility of penta-NiP$_2$ monolayer exhibits remarkable anisotropy, stemming from the anisotropic electronic band structure in the proximity of the Fermi level. Along the [110] direction, the mobilities of the penta-NiP$_2$ monolayer are about 2×10$^3$ cm$^2$V$^{-1}$s$^{-1}$ (electron) and ~10 cm$^2$V$^{-1}$s$^{-1}$ (hole), respectively, which are lower than those along the [100] direction. In addition, we calculated the mobilities of the light carriers along the [110] direction using HSE06 functional, corresponding to the conduction (light electron) and valence (light hole) bands next-nearest to the Fermi level. They are determined to be 5.0 ×10$^4$ cm$^2$V$^{-1}$s$^{-1}$ and 1.4×10$^3$ cm$^2$V$^{-1}$s$^{-1}$, respectively, both of which are higher than those of the heavy carriers along this direction, due to the small effective masses, 0.098 m$_0$ (light electron) and 0.107 m$_0$ (light hole).

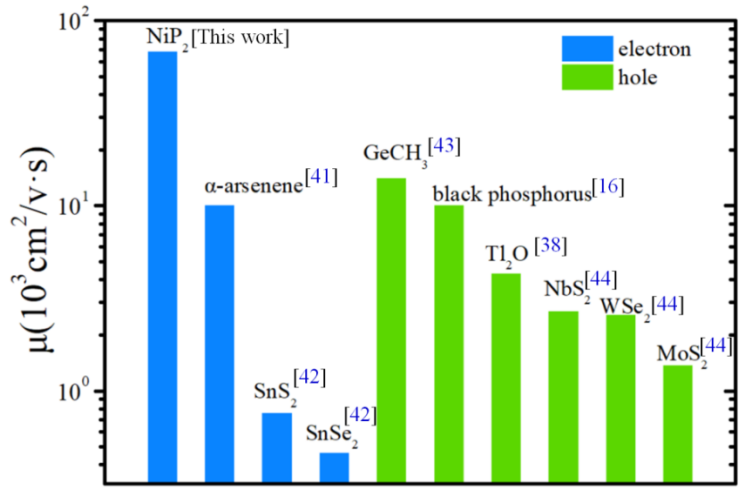

**Figure 5**. Predicted carrier mobilities of typical 2D materials by using an acoustic phonon-limited scattering model on the basis of DFT-HSE06 calculations. Only the highest mobility (electron or hole) of materials is presented for purpose of comparison.

We compared the carrier mobilities of penta-NiP$_2$ monolayer with those of other 2D materials which have been proposed to have high carrier mobility [16,38, 41-44], as shown in Fig. 5. The carrier mobilities of these materials were calculated using the same method, *i.e.* acoustic phonon-limited scattering model in combination with HSE06 functional, to guarantee direct comparison. To strength the comparison, the carrier mobilities obtained by using PBE functional were also presented in Fig. S2 of the Supplementary Information. Clearly, penta-NiP$_2$ monolayer has the highest carrier mobility among these 2D materials. The ultrahigh carrier mobility of penta-NiP$_2$ monolayer that stems from the *p-d* π-conjugation is quite promising for high-speed electronic devices such as field-effect transistors (FET).

### 4. *Stability of penta-NiP$_2$ monolayer.*

NiP$_2$ crystals have two stable phases, mono-clinic (C12/c1) [33,34] and cubic (Pa-3) [35] structures, as shown in Fig. 6(a) and Fig. S3, which have been synthesized in different conditions. The optimized lattice parameters in this work agree well with the experimental data, as listed in Table I, validating the employed theoretical strategy. These two phases are almost energetically degenerate with the cubic structure slightly more stable than the monoclinic structure by about 18.6 meV/atom. In monoclinic NiP$_2$ crystal, we noticed the buckled pentagonal lattice features of the $(10\bar{1})$ planes which are connected by P-P bonds, as shown in Fig. 6(a) and 6(b). A penta-NiP$_2$ monolayer can be obtained by breaking these P-P bonds. In experiments, this may be achieved by growing a NiP$_2$ monolayer on suitable substrate materials, in analogous to the relation between silicene and bulk silicon crystal [45, 46]. The energy of penta-NiP$_2$ monolayer relative to monoclinic NiP$_2$ crystal is 0.543 eV/atom. For comparison, we calculated the formation energy of silicene relative to bulk silicon crystal and found it to be about 0.744 eV/atom. This value is higher than that of penta-NiP$_2$ monolayer by about 40%. Silicene has been successfully grown on different substrate materials. The realization of penta-NiP$_2$ monolayer is therefore expectable in experiments.

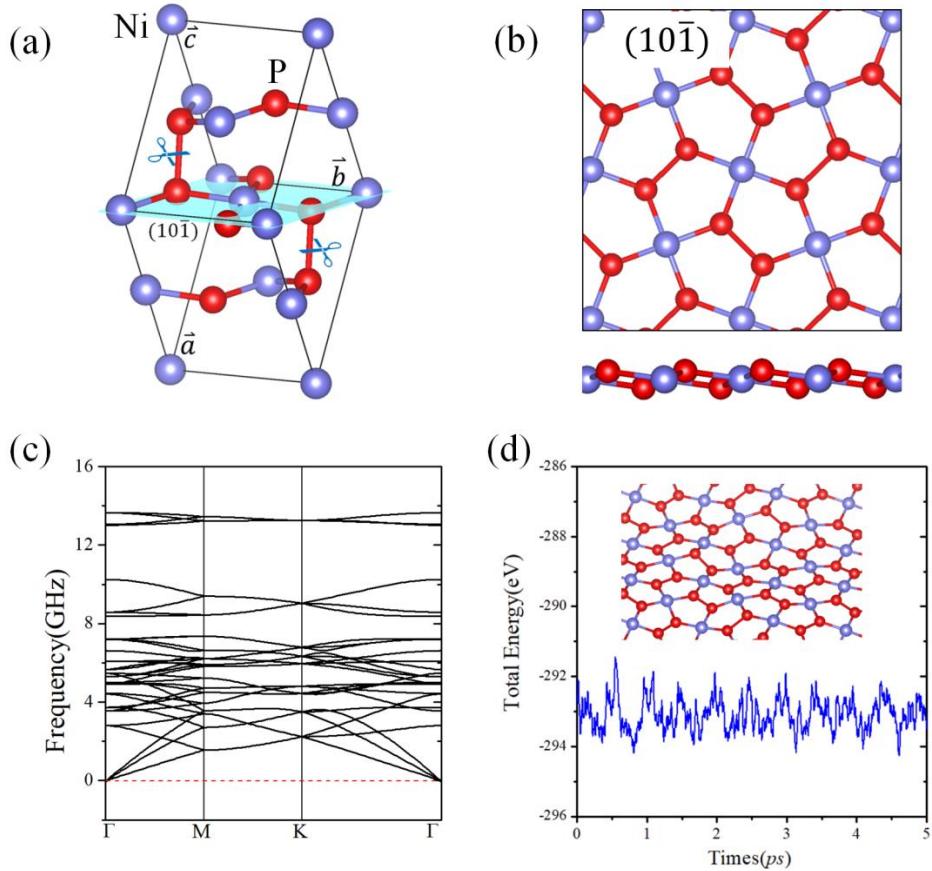

**Figure 6** (a) Crystal structure of monoclinic NiP$_2$ crystal. Ni and P atoms are represented by blue and red balls. (b) Top and side views of the atomic arrangement on the $(10\bar{1})$ plane of monoclinic NiP$_2$ crystal. (c) Phonon spectrum of penta-NiP$_2$ monolayer. (d) Molecular dynamics simulations of penta-NiP$_2$ monolayer at temperature of 500K for 5 ps. The configuration at the end of the simulation is plotted in the inset of this figure which shows the small ripple of penta-NiP$_2$ monolayer at high temperature.

**Table I**. Structural parameters and cohesive energies of monoclinic and cubic $NiP_2$ crystals and 2D penta-$NiP_2$ monolayer. The parameters $d_{Ni-P}$, and $d_{P-P}$ are respectively the lengths of the Ni-P and P-P bonds. The unit of lengths is in angstroms (Å) and that of energies is in eV/atom.

| Structure | Space group | Lattice vectors (Å) | $d_{Ni-P}$ (Å) | $d_{P-P}$ (Å) | Energy (eV/atom) |
|---|---|---|---|---|---|
| Monoclinic | C12/c1 | Theo. a=6.430, b=5.582, c=6.074, α=γ=90°, β=126.77° Exp. a=6.336, b=5.615, c=6.072, α=γ=90°, β=119.62° [33,34] | 2.196 | 2.225 | 0.019 |
| Cubic | Pa-3 | Theo. a=b=c=5.429 Exp. a=b=c=5.471 [35] | 2.266 | 2.190 | 0.00 |
| 2D Pentagonal | P4g | Theo. a=b = 5.552 | 2.167 | 2.091 | 0.562 |

**Table II**. Predicted carrier mobilities of penta-$NiP_2$ monolayer and the relevant parameters along the [100] and [110] directions. $m_0$ is the mass of a free electron. The unit of carrier mobility is $cm^2V^{-1}s^{-1}$.

| Carrier Type | [100] | | | | [110] | | | |
|---|---|---|---|---|---|---|---|---|
| | $m^*/m_0$ | $E_l/eV$ | $C_{2D}/J·m^{-2}$ | $\mu×10^3$ | $m^*/m_0$ | $E_l/eV$ | $C_{2D}/J·m^{-2}$ | $\mu×10^3$ |
| | PBE | | | | | | | |
| e | 0.030 | 1.929 | 123.837 | 200.985 | 0.459 | 2.857 | 115.283 | 5.577 |
| h | 0.031 | 0.562 | 123.837 | 988.861 | 2.334 | 1.307 | 115.283 | 2.281 |
| | HSE06 | | | | | | | |
| e | 0.164 | 1.146 | 111.756 | 39.473 | 0.481 | 2.311 | 119.402 | 3.524 |
| h | 0.204 | 2.964 | 111.756 | 1.988 | 2.203 | 12.725 | 119.402 | 0.011 |

To verify the stability of penta-$NiP_2$ monolayer, we calculated the phonon spectrum by using first-principles calculations in combination with a finite displacement method. A $2\sqrt{2} ×$ $2\sqrt{2}$ supercell was employed with a force convergence criteria of 0.001 eV/Å. The phonon spectrum is free from imaginary frequency modes even in the long-wavelength region, as shown in Fig. 6(c), confirming the dynamic stability. We also preformed *ab initio* molecular dynamics simulations (AIMDS) at temperature of 500K for 5 *ps* to examine its thermal stability, as shown in Fig. 6(d). The total energy of the system converged to a stable value without any tendency of structural decomposition in this time scale. This suggests that penta-NiP$_2$ monolayer can stably exist at or below this temperature once synthesized. The elastic constants of penta-NiP$_2$ monolayer determined from the strain energy in response to in-plane lattice distortion [7, 47] are listed in Table III. Clearly, the mechanical stability criterion for a 2D material, $C_{11} \times C_{22} - C_{12}^2 > 0$ and $C_{66} > 0$, are satisfied. The Young modulus along the x- and y-directions is about 1/3 of that of graphene (335 GPa•nm) [48] and comparable to that of MoS$_2$ (123 GPa•nm) [49]. A positive Possion's ratio is obtained for the planar configuration.

**Table III.** Mechanical and electronic properties of 2D penta-NiP$_2$ monolayer.

| Elastic constants (GPa·nm) | Young's moduli (GPa·nm) | Possion's ratio | Band Gap (eV) |
|---|---|---|---|
| $C_{11} = C_{22} = 123.89$; $C_{12} = 28.33$; $C_{66} = 37.78$ | 117.42 | 0.229 | 0.069 (PBE) 0.818 (HSE06) |

**IV. CONCLUSION**

In summary, we propose a tight-bonding model of a two-dimensional Cairo pentagonal lattice and demonstrate that the *p-d* π-conjugation in this unique framework leads to intriguing electronic properties, such as intrinsic direct band gap, ultra-high carrier mobility and slant Dirac cones. Using DFT calculations, we predict a novel two-dimensional material, penta-MX$_2$ monolayer with M = Ni, Pt and Pd, X = N, P, As and Sb, to realize the predictions of the TB model. It is found that penta-NiP$_2$ monolayer an intrinsic direct band gap of 0.818 eV and ultra-high carrier mobility (~$10^5$-$10^6$ cm$^2$V$^{-1}$s$^{-1}$) comparable to that of graphene. The stability and possible synthetic routes of penta-NiP$_2$ monolayer are verified from different aspects. The π-conjugated Cairo pentagonal lattice with diverse electronic band structures revealed by both TB model and DFT calculations not only enriches the family of 2D pentagonal materials but also find applications in high-speed electronic devices.

## ASSOCIATED CONTENT

### Supporting Information

Electronic band structure of penta-MX$_2$ monolayer; Carrier mobilities of 2D materials predicted at the PBE level; Crystal structure of cubic NiP$_2$ and 2D penta-NiP$_2$ monolayer

## ACKNOWLEDGMENT

MWZ acknowledges financial support from the National Key Research and Development Program of China (Grant No. 2016YFA0301200), the National Natural Science Foundation of China (Nos. 21433006 and 11774201) and the 111 project (No. B13029).